\documentclass[preprint,12pt]{elsarticle}
\usepackage{amssymb}
\usepackage{amsmath}
\usepackage{braket}
\usepackage{graphicx}
\usepackage{dcolumn}
\usepackage{bm}
\usepackage{booktabs}
\usepackage{float}
\usepackage{color}
\usepackage{xcolor}
\usepackage[colorlinks=true,citecolor=blue,linkcolor=blue,urlcolor=blue]{hyperref}
\usepackage{hyperref}
\journal{Computer Physics Communications}

\begin{document}

\begin{frontmatter}
\title{Assessing the Advantages and Limitations of Quantum Neural Networks in Regression Tasks}%

\author[label1]{ Gubio G. de Lima}\ead{gubiofisika@gmail.com}
\author[label1]{Tiago de S. Farias}%
\author[label1]{ Alexandre C. Ricardo}%
\author[label1]{Celso Jorge Villa Boas}

\affiliation[label1]{organization={Universidade Federal de São Carlos},
            addressline={Rodovia Washington Luís, km 235 - SP-310},
            city={ São Carlos},
            postcode={13565-905},
            state={SP},
            country={Brazil}}

\begin{abstract}
The development of quantum neural networks (QNNs) has attracted considerable attention due to their potential to surpass classical models in certain machine learning tasks. Nonetheless, it remains unclear under which conditions QNNs provide concrete benefits over classical neural networks (CNNs). This study addresses this question by performing both qualitative and quantitative analyses of classical and quantum models applied to regression problems, using two target functions with contrasting properties. Additionally, the work explores the methodological difficulties inherent in making fair comparisons between QNNs and CNNs. The findings reveal a distinct advantage of QNNs in a specific quantum machine learning context. In particular, QNNs excelled at approximating the sinusoidal function, achieving errors up to seven orders of magnitude lower than their classical counterparts. However, their performance was limited in other cases, emphasizing that QNNs are highly effective for certain tasks but not universally superior. These results reinforce the principles of the ``No Free Lunch'' theorem, highlighting that no single model outperforms all others across every problem domain.
\end{abstract}

\begin{keyword}
Quantum neural network \sep comparative study \sep regression tasks.
\end{keyword}

\end{frontmatter}


\section{Introduction}

Classical artificial neural networks have become a cornerstone of modern machine learning and are applied across numerous scientific and engineering domains \cite{Schalkoff,pedrycz2020deep,aplication1,aplication2,aplication3,aplication4,aplication5,aplication6}. Their power stems from their proven ability as universal function approximators \cite{aproximator1,aproximator2,aproximator3}, which, when combined with sophisticated optimization techniques \cite{otimization1,otimization2,otimization3}, allows them to model complex, non-linear relationships directly from data \cite{datadrive1,datadrive2}. Despite their widespread success, significant challenges remain. These include the ``black box'' nature of deep models, which hinders interpretability, the need for substantial labeled datasets, catastrophic forgetting, vanishing, and exploding gradient problems, and the escalating computational and energy costs of state-of-the-art architectures \cite{challenger1,challenger2}. The latter presents a significant barrier for applications in resource-constrained environments and for researchers without access to large-scale computing infrastructure.

In this context, quantum neural networks have emerged as a promising alternative, attempting to leverage properties of quantum mechanics to overcome the limitations of classical models in certain tasks \cite{QNN0,QNN1,QNN2,QNN3}. Nevertheless, the exact quantum properties that may contribute to these improvements, as well as the nature of the tasks in which such advantages manifest, remain open questions in the scientific community. Notwithstanding the interest in QNNs, their definition and practical benefits are still under active debate \cite{QNN4,QNN5,QNN6}. A particularly relevant approach involves the use of continuous-variable quantum computing as a model of execution of QNNs  \cite{CV-QNN1}. This model presents significant advantages compared to the qubit or qudit-based quantum computing paradigms, such as the ease of encoding floating-point numbers and the inherent inclusion of nonlinear operations within quantum circuits \cite{CV-QNN2,CV-QNN3,CV-QNN4,CV-QNN5,CV-QNN6,CV-QNN7}.

Several studies have focused on comparing classical and quantum algorithms \cite{comparasion2,comparasion3,comparasion5,comparasion6,comparasion7,comparasion9,comparasion10,comparasion11,comparasion12,comparasion13}, with different conclusions and perspectives. Abbas et al. \cite{QNN0} evaluated the expressibility and training efficiency of a QNN, employing information geometry to define the expressibility of quantum versus classical models, providing evidence that well-designed QNNs can surpass classical neural networks in terms of effective dimension and training speed, with validation on real quantum hardware. In the work of Qian et al. \cite{QNN7} they conducted numerical experiments and observed that, although QNNs can perform well on specific datasets, they generally fail to outperform classical models across broader tasks, indicating that the practical benefits of QNN are still under investigation and not yet universally established.
Among these studies, only a limited number investigate CNNs and QNNs \cite{comparasion1,comparasion4,comparasion14}. Within this subset, the majority of applications focus on classification problems, whereas only a few address regression-like tasks \cite{comparasion5,comparasion6,comparasion7}.
Regression constitutes a central problem in machine learning, as it involves estimating feature mappings to real-valued outcomes, thereby enabling quantitative predictions that directly inform decision-making. Numerous real-world applications rely on regression analysis, including forecasting \cite{regression2}, control \cite{regression1}, energy systems \cite{regression4}, finance \cite{regression5}, and scientific research \cite{regression6}

Establishing a fair comparison between QNNs and their classical counterparts is notoriously difficult, not only due to fundamental differences in hardware architectures, but also because of the absence of universally accepted criteria for such an analysis \cite{comparasion1,comparasion2,comparasion3,comparasion4}. This challenge extends beyond the fundamental differences, as even within classical machine learning, comparing models with disparate architectures and hyperparameter spaces lacks a standardized methodology.

\begin{figure*}
    \centering
    \includegraphics[width=1\linewidth]{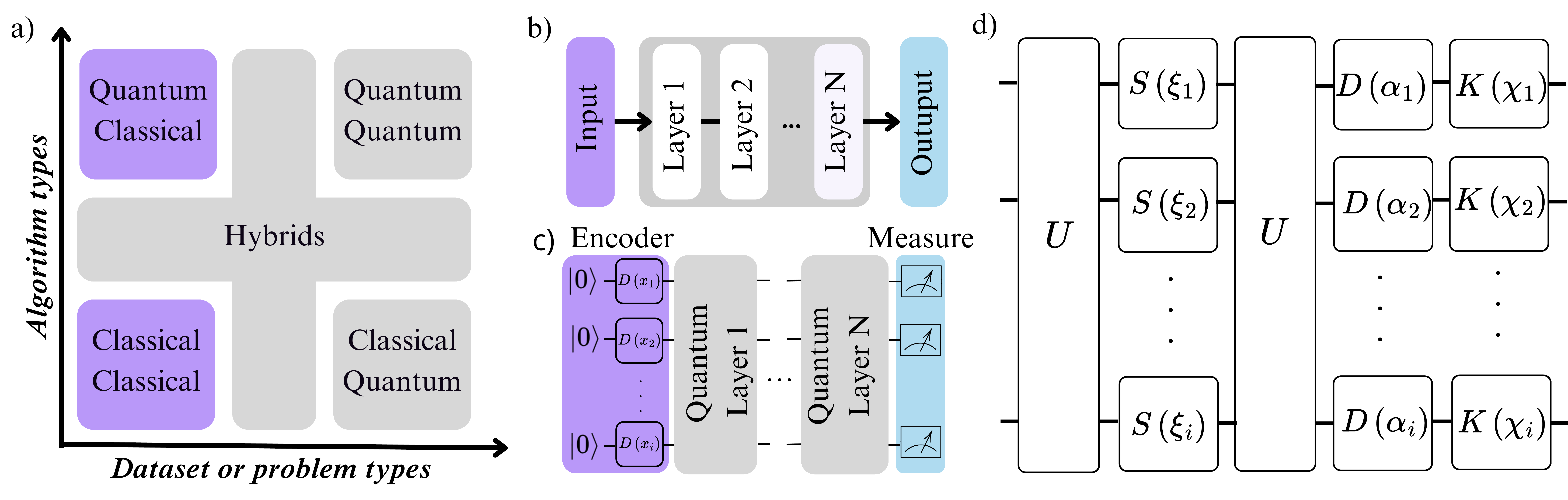}
    \caption{ (a) Algorithm versus problems addressed by quantum and classical ML; cases highlighted in purple represent those explored in this work. (b) Schematic representation of the neural network model. (c) Representation of the variational quantum algorithm (QNN), with each line representing a quantum mode. (d) Detailed representation of the ``Quantum layer'' described in (c), illustrating the sequence of quantum logic gates employed. }
    \label{fig:primeira}
\end{figure*}

The domain of quantum machine learning (QML) encompasses a wide landscape of potential algorithm and data-type combinations, as illustrated in Figure \ref{fig:primeira}-a). This work situates itself within the hybrid quantum-classical regime, where classical problems are tackled using both quantum and classical algorithms. To benchmark the performance of CNNs and QNNs, we select regression tasks involving two distinct target functions: a smooth, continuous sinusoidal function and a discontinuous Heaviside step function. Our analysis is conducted using two strategies: first, varying the number of layers, and second, matching the number of trainable parameters across models. We evaluate prediction accuracy using the Mean Squared Error (MSE) as a quantitative metric. The primary objective is to identify specific scenarios in which QNNs demonstrate a concrete advantage over CNNs, while also discussing the inherent challenges and limitations of such a comparative analysis.

The article is structured as follows. In Section~\ref{secao2}, we present the core component of the classical neural network algorithm: the neuron, or perceptron. Section~\ref{secao3} introduces the continuous-variable quantum neuron used in the comparison, along with the encoding process and measurement scheme. In Section~\ref{secao4}, we discuss the challenges and complexities involved in developing a fair and meaningful comparison between the algorithms, given their intrinsic differences. Section~\ref{secao5} presents and analyzes the main results of the comparative study. Finally, in Section~\ref{secao6}, we summarize the conclusions and outline future research directions in this area.

\section{Classical Neural Networks}\label{secao2}

A neural network is a computational model composed of interconnected units known as artificial neurons, here named as classical neurons, organized in different layers. The mathematical function defines the $k$-th classical neuron as 
\begin{equation}
    G_{k}^{[l]} \left( \sum_i w_{i,k}^{[l]} \cdot x_{i}^{[l-1]} + b_{k}^{[l]} \right),
    \label{classicalneuron}
\end{equation}
which depends on the trainable parameters $ w_{i,k}^{[l]} $  and  $b_{k}^{[l]}$, known as weights and biases, for the $k$-th neuron in the $l$-th layer. These parameters determine the response of the neuron to the input data $ x_{i}^{[l-1]} $ from the preceding layer. The function $G_{k}^{[l]} $ represents the non-linear activation function. A neural network is formed by organizing these neurons into layers, where their collective interactions process information.

A common neural network architecture, particularly in supervised learning, is the fully connected feedforward neural network \cite{CNN1}. In this architecture, each layer processes the information and passes it to the next layer, as illustrated in Figure~\ref{fig:primeira}-b).
The performance and versatility of this model stem not only from its structural design, but also from the learning algorithms that allow its parameters to be efficiently updated.  Training is typically achieved by minimizing a loss function using gradient-based optimization algorithms \cite{Soydaner-otimization, FatimaNN-optimization}. The required gradients are efficiently calculated via the backpropagation algorithm \cite{brackpropagationNN}, which is implemented in modern software frameworks using automatic differentiation techniques \cite{CNNautomaticdiff,CNNautomaticdiff2}.

According to the universal approximation theorem \cite{CNNteoremaU}, a neural network can approximate any continuous function, provided it has at least one hidden layer and a sufficient number of neurons. 
However, the theorem is non-constructive, since it does not specify the required number of neurons, layers, or the optimal choice of activation function. These architectural choices thus become critical hyperparameters. Consequently, designing an effective neural network for a specific problem often requires extensive hyperparameter tuning or an architecture search process to find a suitable configuration.

\section{Quantum Neural Networks}\label{secao3}

The exact definition of a quantum neural network is an ongoing debate in the scientific community, with proposals ranging from models that integrate classical and quantum elements to purely quantum implementations, reflecting the emerging nature of this field of research \cite{QNN1, ezhov2021on,comparasion1,abel2022completely,zhou2023quantum}. A common framework, and the one adopted here, is the variational quantum algorithm, which user parameterized quantum circuits as trainable models analogous to classical networks.

In this work, we employ a continuous-variable (CV) QNN based on the variational circuit model introduced in Ref. \cite{CV-QNN1}. This paradigm is particularly well-suited for our comparative analysis due to two key advantages over qubit-based approaches: the natural encoding of real-valued data and the ability to implement non-linear operations directly within the quantum circuit. Our implementation, shown schematically in Figures 1-c) and 1-d), operates in three stages: data encoding, parameterized processing, and measurement.

\begin{enumerate}
    \item \textbf{Data encoding:} To input a classical value $x$ into the network, we apply a displacement operator $\hat{D}(x)$ to the vacuum state $|0\rangle$, preparing the initial state $|x\rangle = \hat{D}(x) |0\rangle$.  This method directly maps a real number to the phase space of a quantum mode, avoiding the significant overhead required to represent continuous data in qubit-based  \cite{QNNencoder,QNNencoder2,QNNencoder3}.

    \item \textbf{Parameterized Quantum Layers:} Following encoding, the state is processed by a Quantum Layer. For a single-mode system, this layer is composed of a sequence of quantum gates that we define as a quantum neuron:
    \begin{equation}
    \textbf{Quantum Neuron}: \hat{K}(\chi)\hat{D}(\alpha)\hat{R}(\theta_2)\hat{S}(\xi)\hat{R}(\theta_1).
    \end{equation}
    This sequence consists of the gaussian operation as displacement $\hat{D}(\alpha)$, rotation $\hat{R}(\theta)$, squeezing  $\hat{S}(\xi)$ \cite{reviewagates}.A critical component for expressive power is a nonlinear transformation. In this case, we user a validated approximation of Kerr gate $\hat{K}(\chi)$ tailored for trapped-ion systems \cite{alexandreQNN}. A deep QNN is constructed by repeating this entire quantum layer $L$ times, as depicted in Figure 1 c). Each layer adds five trainable parameters $(\chi, \alpha, \theta_2, \xi, \theta_1)$, resulting in a total of $5L$ parameters for the model.

    \item \textbf{Measurement:} To extract a classical output from the circuit, a measurement is performed. For the tasks in this study, we measure the expectation value of the quadrature operator $\hat{X} = \frac{1}{\sqrt{2}}(\hat{a}^\dagger+\hat{a})$, where $\hat{a}$ and $\hat{a}^\dagger$ denote the annihilation and creation operators. This approach makes it easier to obtain a real-valued floating-point number in the output.
\end{enumerate}

Our analysis focuses on single-variable problems, restricting the QNN implementation to a single quantum mode. Note that, in this case $\hat{U}$ in Figure 1-d) is given by the rotation operator. This decision is pragmatic, as simulating multi-mode CV systems on classical hardware is computationally demanding due to the exponential growth of the Hilbert space. For the numerical simulation, quantum states are represented in the Fock basis, which consists of photon number states $|n\rangle$. Any pure state $|\psi\rangle$ is expanded as an arbitrary superposition of the Fock states as $|\psi\rangle = \sum_n c_n |n\rangle$. Because the full basis is infinite-dimensional, we must truncate it at a finite cutoff dimension for numerical calculations. This is a valid approximation provided that the probability amplitudes $c_n$ for $n$ near the cutoff are negligible. In this work, we set the cutoff dimension to 30, which ensures simulation fidelity while maintaining computational tractability.

\section{Challenges in Comparison}\label{secao4}

Establishing a fair comparison between classical and quantum neural networks is intrinsically complex. The challenge stems not only from their disparate computational architectures but also from the absence of universally accepted benchmarking criteria. The ``No Free Lunch'' theorem \cite{nofreelacnh3} underscores this, asserting that no single algorithm is optimal across all possible problems. An algorithm superior in one problem class may be suboptimal in another \cite{nofreelacnh1,nofreelacnh2}. Therefore, any meaningful comparison requires a clearly defined context, including a specific problem scope, suitable evaluation metrics, and a careful accounting of structural differences between the models.

To create this well-defined context, we first narrow our focus to supervised regression tasks. This choice is motivated by several factors. First, supervised learning is a foundational paradigm in both classical and quantum machine learning, providing common ground for comparison. Second, for single-variable regression, estimating an unknown function $f(x)$ from input-output pairs $\{(x^{(i)}, y^{(i)})\}_{i=1}^n$, aligns directly with the one-input, one-output architecture of the continuous-variable QNN model used in this work. The goal is to train a model $\hat{f}$ that accurately predicts an output $\hat{y} = \hat{f}(x)$ for a new input $x$.

Next, we must select an evaluation metric. Performance can be measured from multiple perspectives, including accuracy, training time, computational cost, or robustness. Since our primary goal is to evaluate the expressive power and solution quality of the models, we adopt the MSE as both the training loss function and the final performance metric. The MSE is defined as
\begin{equation}
    \text{MSE} = \frac{1}{N} \sum_{i=1}^{N} (\hat{y}_i - y_i)^2,
\label{MSE}
\end{equation}
where $N$ is the number of data points, $y_i$ is the true value, and $\hat{y}_i$ is the prediction of the model. Finally, to navigate the structural disparities between the models, we devise two complementary comparison strategies. Each strategy is designed to isolate different aspects of performance and provide a nuanced view of the capabilities of each model.

The first approach involves using one classical neuron and one quantum neuron per layer while keeping as many hyperparameters as possible fixed, varying only the number of layers. This design ensures that the network architectures remain as comparable as possible. However, this approach has inherent limitations. 
In particular, it raises questions about the validity of the comparison since classical networks offer a variety of activation functions, whereas QNNs rely on a specific nonlinear function, controlled by the parameter $\chi$.  Moreover, each QNN layer includes five trainable parameters, whereas the simplified classical neuron has only two. 
This gives the QNN a greater representational capacity for the same number of layers. This structure artificially limits the classical model, which typically leverages wider layers (more neurons per layer) to achieve universal approximation. This discrepancy raises concerns about the validity of comparing models that differ in their parameter counts, given that the number of parameters influences the optimizer and, consequently, the algorithm's accuracy and precision.

Our second strategy aims to provide a fairer comparison of expressive power by ensuring both models have the same number of trainable parameters. To achieve this, we fix the QNN architecture (and thus its parameter count) and adjust the width and depth of the classical network to match that count. Although this equalizes model capacity, it reveals a fundamental architectural trade-off: in classical networks, parameters can be flexibly distributed between width and depth, whereas in our QNN the parameterization is strictly determined by its depth (number of layers). This approach allows for a more direct evaluation of which model makes more effective use of its parametric budget.

\section{Results and Discussion}\label{secao5}

In this section,  we present a comparative analysis of the classical and quantum neural networks on supervised regression tasks.  To evaluate the models' performance, we selected two target functions with distinct properties over the interval $[-1,1]$. The first is the sine function $ y = \sin(\pi x) $ , a common choice for regression tasks due to its continuous and periodic nature, which appears in various physical and natural phenomena. The second function is the Heaviside function, given by, 
\begin{equation*}
    \text{Heaviside}(x) =
    \begin{cases} 
        0 & \text{if } x < 0, \\
        1 & \text{if } x \geq 0.
    \end{cases}
\end{equation*}

Unlike the sine function, the Heaviside function exhibits a discontinuity at $x = 0$ and is therefore non-differentiable at this point. 
This property imposes an additional challenge on the regression model, as it requires the ability to handle abrupt transitions in the target function values. Both functions are commonly used in testing numerical methods of approximating functions  \cite{resultsRegression,resultsRegression1,resultsRegression2}. 

\subsection{Implementation Details}  

The data used for training and testing were generated through a uniform discretization of the domain of the selected functions. 
A total of $N_{\text{train}} = 20$ points was used for the training set, and, to ensure a robust evaluation of the models, $N_{\text{test}} = 200$ points were used for the test set.

For the optimization strategy, we employed the Adam algorithm \cite{resultsADAM} with a learning rate of 0.01 over $10^4$ epochs in all cases.  To account for the stochasticity of parameter initialization, all reported results are averaged over 100 independent training runs with different random seeds. The choice of learning rate and number of epochs was guided by prior experience with similar models, as well as preliminary tests conducted on the current setup. These exploratory runs indicated that a learning rate of 0.01 provides a good balance between convergence speed and training stability across both models considered. Furthermore, the number of training epochs was constrained by the computational cost of simulating the quantum model.

The simulation time of the entire quantum system for one of the functions took approximately one week, even when parallelized across a machine with two AMD EPYC™ 7452, 32 CPU Cores(128 thread) and 256GB Ram(3200Mz). This limitation imposed a practical upper bound on the number of epochs feasible within our computational budget.

\subsection{Comparison in terms of the Number of Layers}

\begin{figure}
    \centering
    \includegraphics[width=1\linewidth]{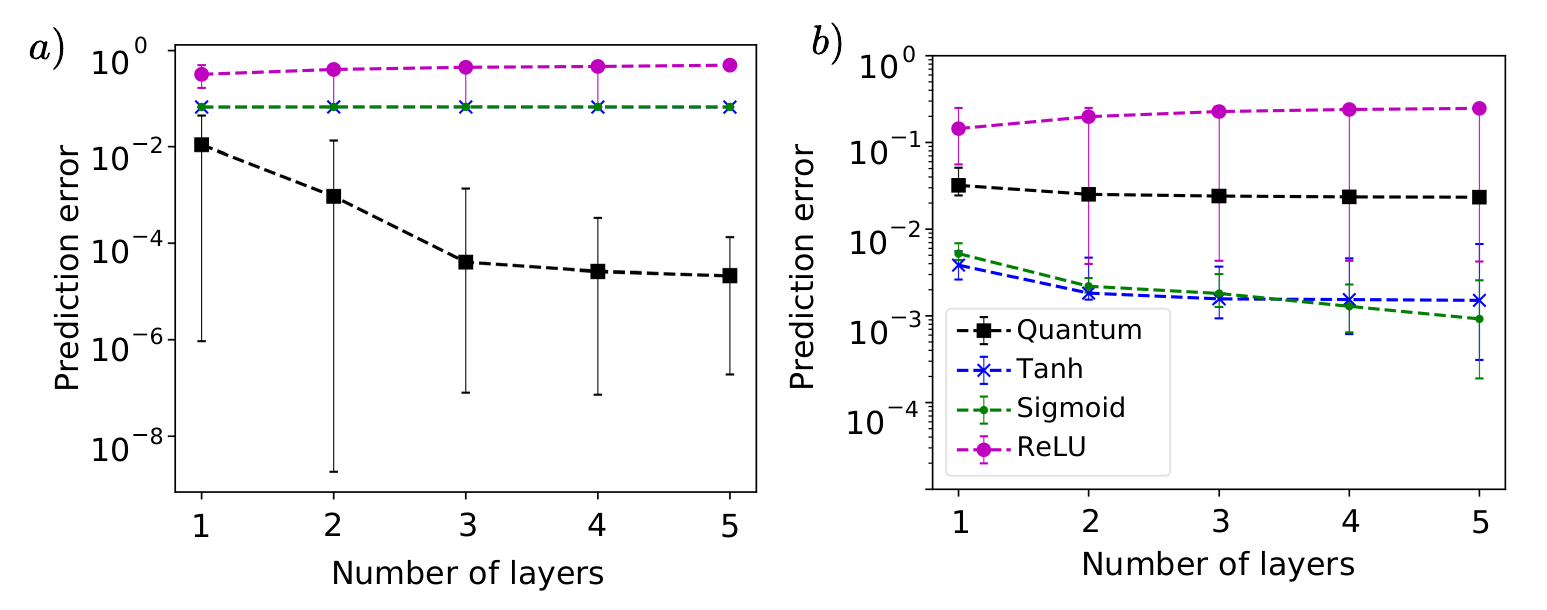}
    \caption{Prediction error with MSE in logarithmic scale of QNN and CNN, as a function of the number of layers. The curves represent different activation functions used in CNN: Tanh (blue), Sigmoid (green), and ReLU (magenta). The black squares indicate results obtained by the quantum model. Error bars represent statistical variation across 100 different initializations per layer. Subplot a) presents the results for the $sin(\pi x)$ and b) the Heaviside(x).}
    \label{fig:figura2}
\end{figure}

We first analyze the models under our structural strategy, where performance is evaluated as a function of network depth with a single neuron per layer. This test is designed to probe how each architecture leverages increasing depth under a highly constrained, narrow structure. As shown in Figure \ref{fig:figura2}-a), the quantum model (black dotted line) exhibits a decreasing trend in error as the number of layers increases. Notably, the quantum algorithm achieved the best performance, reaching an MSE on the order of $10^{-8}$. 
This behavior indicates that the quantum architecture effectively captures the smooth structure of the sine function, reducing the error by several orders of magnitude while increasing network depth. 

In contrast, all classical models stagnate with high error, regardless of depth. This poor performance is an expected consequence of the single neuron per layer constraint. A narrow classical network has severely limited representational capacity, as it cannot form the complex combinations of features that wider layers enable. The QNN, whose parameters are inherently tied to its depth via a sequence of unitary operations, does not suffer from this specific limitation in the same way, highlighting a fundamental difference in how the two architectures user their parameters. Our hypothesis for the observed quantum advantage is attributed to the higher expressiveness associated with the number of parameters. Specifically, in the case of CNNs, the parameter count scales as $2L$, whereas in QNNs it scales as $5L$.

\begin{figure}
    \includegraphics[width=1\linewidth]{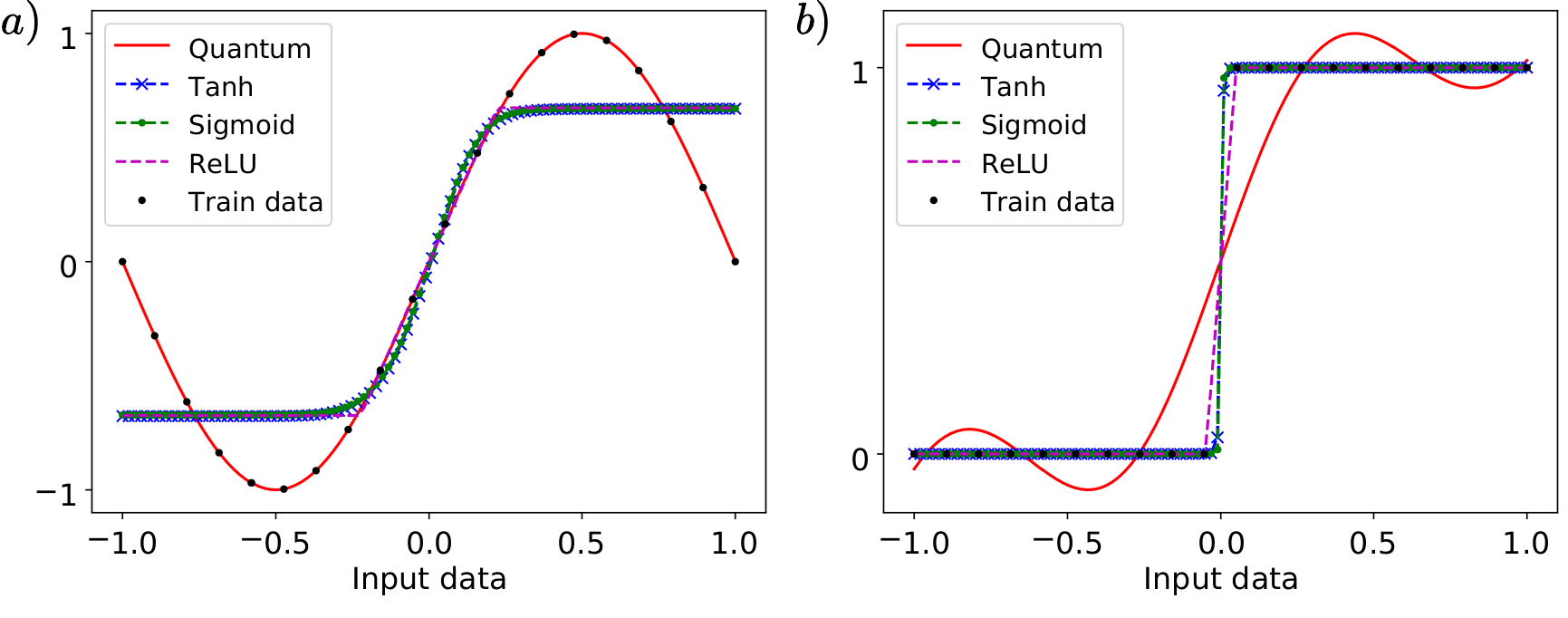}
    \caption{Best results obtained after training the algorithms applied to the (a) sine and (b) Heaviside functions. The curves represent the quantum neural network (red) and classical neural networks with activation functions: Tanh (blue), Sigmoid (green), and ReLU (magenta), each curve was generated with $200$ points. The black points indicate the training data with $20$ points . }
    \label{fig:figura3}
\end{figure}

Figure \ref{fig:figura2}-b) presents the results for approximating the Heaviside function. In this case, a markedly different behavior is observed compared to the previous result. The quantum model maintains a relatively high and nearly constant error as the number of layers increases, suggesting a fundamental limitation in its ability to represent abrupt discontinuities. In contrast, classical models with Tanh and Sigmoid activations show a progressive reduction in error as the number of layers increases. Their inherent sigmoidal shape is well-suited to approximating a step-like transition, and adding depth allows for a steeper and more accurate fit. The ReLU-based model, however, continues to exhibit the worst performance, as its piecewise-linear nature is ill-suited for this task.

Figure \ref{fig:figura3} provides a qualitative comparison of the predictive capabilities of the best results of classical and quantum neural networks in modeling the a) sine and b) Heaviside functions. For the sine function, the quantum model (red curve) exhibits good fitting and generalization in the training region and the classical networks with \(\tanh\), sigmoid, and ReLU activation functions fit the data well in the central region, where the input data ranges over $ [-0.25, 0.25]$, but struggle at the extremes. 
The three classical cases have good results for the Heaviside function, although ReLU, on average, presents poor results compared to the quantum model, as we see in Figure \ref{fig:figura3}-b). The failure of the QNN is also clear, as its prediction is characterized by the smooth, oscillatory behavior intrinsic to its variational form, which prevents it from capturing the sharp transition.

\subsection{Comparison in Terms of the Number of Parameters}  

In the previous results, we identified a case in which the quantum algorithm outperformed the classical one. However, this comparison was limited to CNNs, as the number of parameters in the two models differed. To address this, our second strategy offers an alternative comparison. In this approach, we evaluate the models based on their parameter count. Here, the goal is to assess which architecture makes more efficient use of a given number of trainable parameters. For each parameter count defined by the QNN, we explored multiple CNN architectures with different combinations of layers and neurons that yielded the same total parameter count. The full details of these configurations are provided in the Appendix \ref{tab:config_neural} .

Figure \ref{fig:figura4} presents the comparison of the prediction errors obtained by each algorithm as a function of the number of parameters. The plot in Figure \ref{fig:figura4}-a) displays the MSE for the function \( \sin(\pi x) \) on a logarithmic scale, allowing for a clear visualization of the performance differences among the tested models.  The CNN exhibits relatively high errors, with the ReLU function performing particularly poorly, and the other two activation functions produce results with high variance. On average, it shows worse results. Only in specific situations does the CNN obtain good results. In contrast, the quantum model (represented by black squares) demonstrates a steeper decline in error from  $10^{-2}$ to $10^{-5}$, reaching average values three orders of magnitude smaller than the average of some classical activation functions. The results of the quantum model used in this second comparison are the same as those in the previous subsection, so we will have the same result. And as previously mentioned, they were the reference for defining the number of parameters in the comparison. Notably, the quantum model with only five parameters, shown in Figure \ref{fig:figura4}-a), already achieves results that classical models fail to reach. These findings suggest that, on average, the quantum neural network with continuous variables is capable of capturing the function’s structure more effectively, even with a limited number of parameters. This indicates a potential advantage in learning smooth and continuous functions. 

\begin{figure}
    \includegraphics[width=1\linewidth]{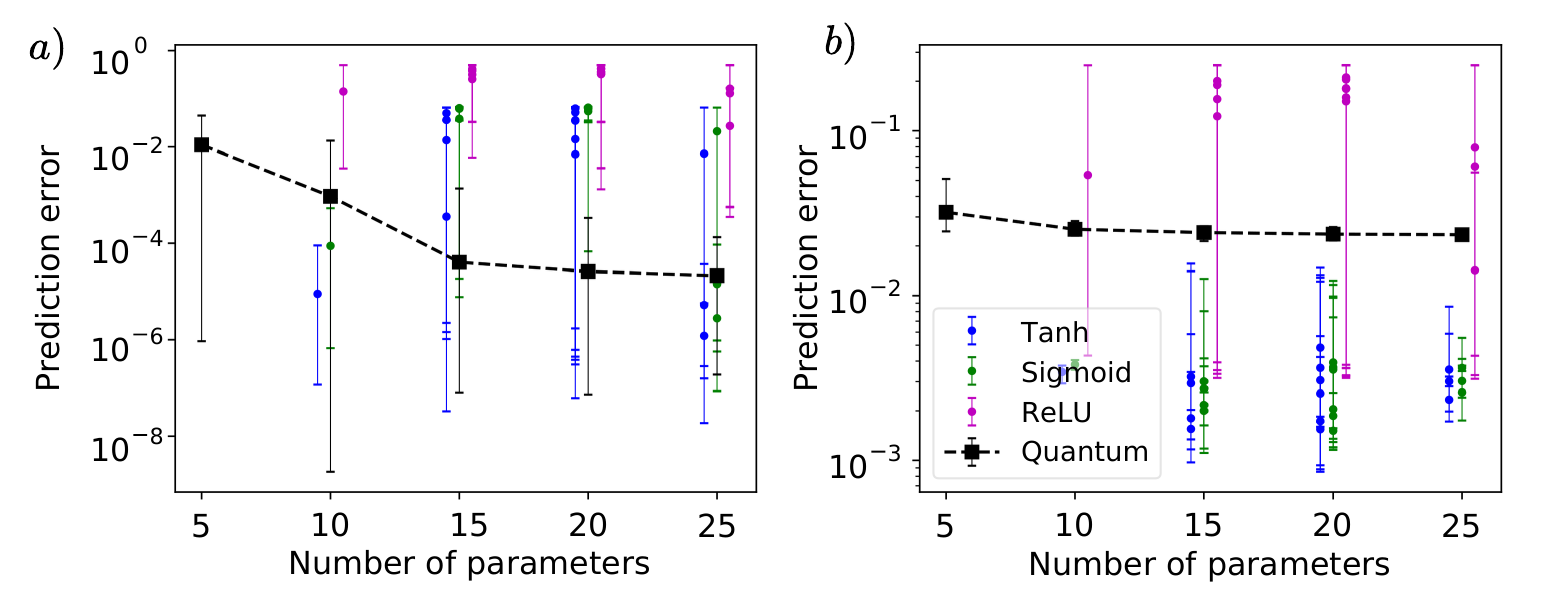}
    \caption{Prediction error results for each algorithm as a function of the number of parameters. The y-axis represents the error with MSE on a logarithmic scale, while the x-axis indicates the number of parameters with different combinations of neurons and layers for each point. The colors represent different activation functions used in CNN: Tanh (blue), Sigmoid (green), and ReLU (magenta). The black squares indicate results obtained by the quantum model. Error bars represent statistical variation across 100 different initializations per layer. Subplot a) presents the results for the $sin(\pi x)$ and b) the Heaviside(x) functions. }
    \label{fig:figura4}
\end{figure}
The second subplot \ref{fig:figura4}-b) presents results for the Heaviside function. Unlike the previous case, CNN exhibits highly activation-dependent behavior. Classical neural networks with tanh and sigmoid activation functions achieve lower and consistent errors, whereas ReLU-based models show a much greater dispersion and significantly higher errors, highlighting difficulties in learning the discontinuity. Meanwhile, the quantum model maintains a relatively constant error without the decreasing trend observed for the sine function. This suggests that, for discontinuous functions, the quantum model may not offer the same advantage observed for smooth functions, indicating that its applicability may strongly depend on the characteristics of the target function. 

\begin{figure}  
    \centering  
    \includegraphics[width=1\linewidth]{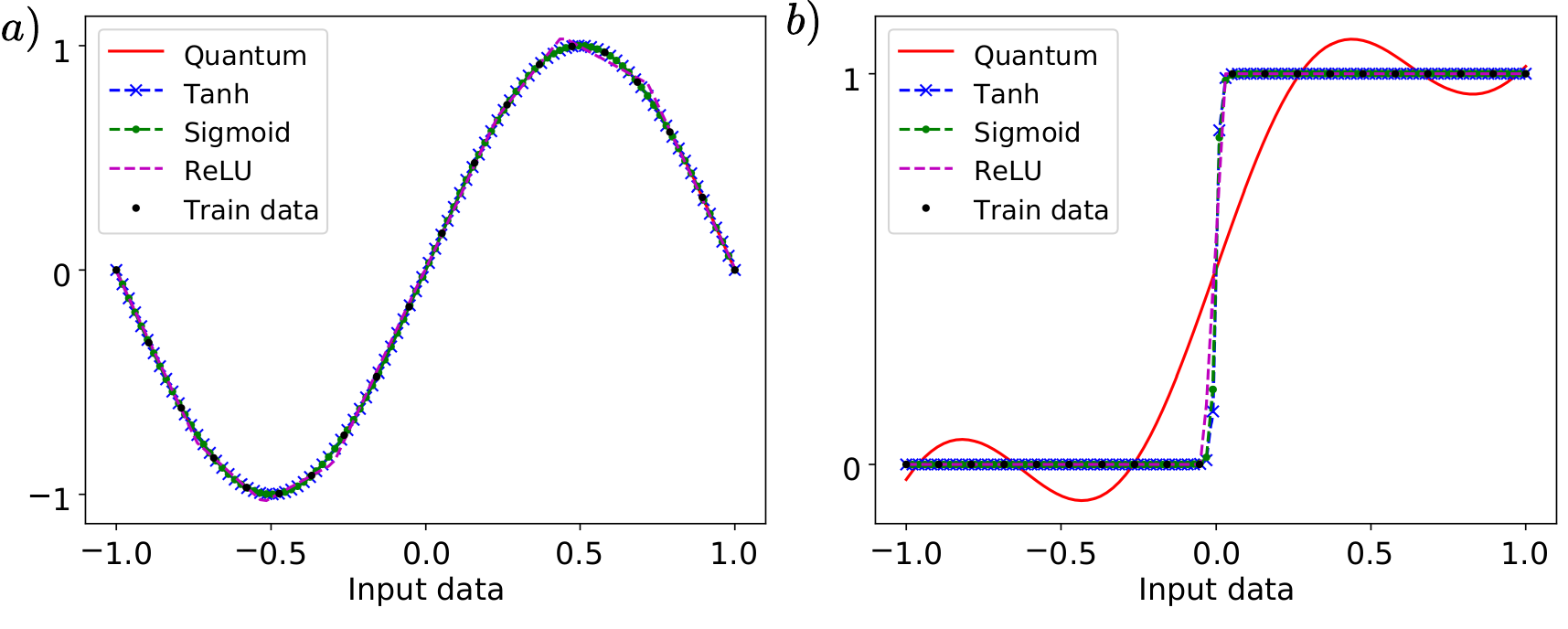}    
    \caption{Best results obtained after training the QNN and CNN  in the comparison based on the number of parameters. Panel (a) presents the fitting results for the sine function, while panel (b) corresponds to the Heaviside function. The curves represent the quantum neural network (red) and classical neural networks with activation functions Tanh (blue), Sigmoid (green), and ReLU (magenta),  each curve was generated with $200$ points. The black points indicate the training data with $20$ points .} 
    \label{fig:figura5}  
\end{figure}

Figure \ref{fig:figura5} provides a qualitative view of the best-performing models under this parameter-matched comparison. The most striking result is seen in Figure \ref{fig:figura5}-a). Unlike the failed attempt in the first strategy (Figure \ref{fig:figura3}-a), the classical networks are now able to perfectly fit the sine function. This starkly illustrates the central theme of our comparative study: the conclusions drawn about the relative strengths of QNNs and CNNs are fundamentally dependent on the chosen basis for comparison. A comparison by structural analogues suggests classical inferiority on the sine task, whereas a comparison by parameter count demonstrates classical competence. This highlights the critical importance of defining a fair and well-motivated methodology when benchmarking quantum against classical machine learning models.

\subsection{Discussions}  

The comparison reveals that the performance of the quantum model is highly sensitive to the nature of the function being learned. 
While for smooth functions, such as the sine function, it exhibits a clear advantage in terms of prediction error, this advantage is not evident for discontinuous functions, such as the Heaviside function. 
This suggests that, although quantum computing can offer significant improvements in certain scenarios, its effectiveness may depend on the specific problem being addressed, as in the classic case.

The advantage observed in the quantum model for smooth functions in this study may be attributed to the model’s ability to effectively represent such functions within its state space. In contrast, the performance with the Heaviside function indicates a limitation in capturing functions with discontinuities. This reinforces the idea that the choice of a learning algorithm should consider the characteristics of the problem and that significant gains in specific scenarios do not imply the algorithm's superiority in a general sense.  

Beyond the performance, our study underscores that the methodology of comparison dictates the outcome. Compared with structural analogues, the constrained architecture of the classical network led to its failure on the sine task, creating the illusion of quantum advantage. However, when comparing by parameter count, the classical network, given sufficient flexibility, proved perfectly capable. This demonstrates that claims of quantum advantage can be misleading if not contextualized within a carefully justified and transparent benchmarking framework. Simply having more parameters per layer, as our QNN did in the first strategy, can easily be mistaken for a more fundamental algorithmic superiority.

This work highlights that the path to practical quantum advantage in machine learning will likely not involve a single QNN algorithm that outperforms all classical models. Instead, progress will come from identifying specific problem domains whose structure aligns with the inductive bias of a particular quantum algorithm. Furthermore, the selection of hyperparameter, both for the model architecture and the optimizer, remains as crucial in QML as it is in classical ML \cite{Ilemobayo2024hyperparameter, anand_black_2020}. The need for expert-driven tuning and architecture search is not eliminated by the quantum paradigm but rather adapted to a new set of operations and constraints.

\section{Conclusion}\label{secao6}

This study presented a systematic comparison between classical and quantum neural networks in regression tasks, highlighting the conditions under which each model excels.  We find a quantum advantage in a specific quantum machine learning scenario. However, the results demonstrate that the performance of QNNs is highly dependent on the nature of the target function: while they exhibited superiority in approximating the sine function, achieving a mean square error of the order of $10^{-8}$, their performance was limited in modeling the Heaviside function, which required capturing an abrupt discontinuity.  
In contrast, CNNs with tanh and sigmoid activation functions showed greater adaptability to discontinuities, although they were less efficient in approximating smooth functions.  

The analysis also revealed methodological challenges in comparing these models, such as disparities in the number of parameters and the architectural rigidity of QNNs. Additionally, computational limitations in simulating QNNs on classical hardware restricted the exploration of more complex scenarios (multidimensional regression problems). Therefore, until suitable quantum hardware for this algorithm becomes available, its full potential remains unexplored. Consequently, determining whether quantum or classical neural networks perform better remains an open and challenging problem, raising important questions about the criteria for fair algorithmic comparison. In this work, we presented a way to proceed in this context. This highlights the importance of studies where practical comparative analyses are conducted to enrich the quantum machine learning literature, in a manner analogous to what has already been established for classical machine learning algorithms, and reinforces the importance of well-defined criteria for adopting QNNs. 

Promising avenues for future research involve extending QNNs to higher-dimensional problems (with 2 or more quantum modes) and devising novel quantum models tailored to address discontinuous functions. The integration of hybrid (classical-quantum) techniques also emerges as a promising area for leveraging the advantages of both paradigms. 

\section{Acknowledgments}
This work was supported by the Coordenação de Aperfeiçoamento de Pessoal de  Nível Superior (CAPES) - Finance Code 001 and São Paulo Research Foundation (FAPESP) grants No. 2022/00209-6, 2023/14831-3, and 2023/15739-3. C.J.V.-B. is also grateful for the support from the National Council for Scientific and Technological Development (CNPq) Grant No. 311612/2021-0. This work is also part of the CNPq Grants No. 140467/2022-0 and No. 140468/2022-6. 

\appendix
\section{Table}

Table 1 below presents all configurations used in the classical neural network during the comparison by the number of parameters. Each combination represents the distribution of neurons per hidden layer, where the values in parentheses indicate the number of neurons in each layer. The input and output layers were omitted, as they always consist of a single input and a single output.

\begin{table}[h]
    \centering
    \begin{tabular}{c|c}
        \toprule
        \textbf{Number of} \\ \textbf{Parameters} & \textbf{Combinations} \\
        \midrule
        10  & (3) \\
        15  & (1,4) \quad (4,1) \quad (1,2,2) \quad (2,2,1) \\
        20  & (1,1,5)  (1,5,1)  (2,1,4)  (3,1,3)  (4,1,2)  (5,1,1) \\
        25  & (8) \quad (2,5) \quad (5,2) \\
        \bottomrule
    \end{tabular}
    \caption{Neural network configurations per hidden layer in the classical neural network.}
    \label{tab:config_neural}
\end{table}  


\end{document}